\documentstyle[prl,twocolumn,aps,epsf]{revtex}
\begin{document}
\newcommand{\be}{\begin{equation}}
\newcommand{\ee}{\end{equation}}
\newcommand{\bc}{\begin{center}}
\newcommand{\ec}{\end{center}}
\newcommand{\bi}{\begin{itemize}}
\newcommand{\ei}{\end{itemize}}
\newcommand{\bo}{\={o} }
\newcommand{\th}{$\Theta$}

\title{Identifying the protein folding nucleus using molecular dynamics}

\date{\today}
\maketitle
\bigskip
\bigskip

{\bf \noindent Molecular dynamics simulations of folding in an
off-lattice protein model reveal a nucleation scenario, in which a few
well-defined contacts are formed with high probability in the transition
state ensemble of conformations. Their appearance determines folding
cooperativity and drives the model protein into its folded
conformation.}

\begin{figure}[htb]
\centerline{
\epsfxsize=8.0cm
\epsfbox{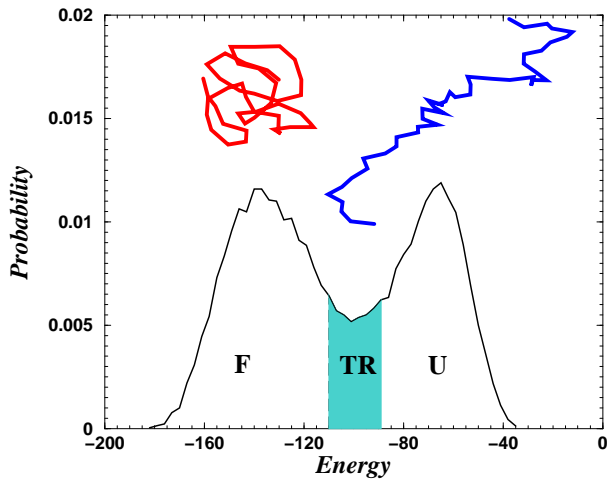} 
\vspace*{1.0cm}
}
\caption{The probability distribution of the energy states $E$ of the
46-mer maintained at the folding transition temperature $T_f=1.44$.  The
bimodal distribution indicates the presence of two dominant states: the
folded (region $F$) and the unfolded (region $U$) states. The transition
state ensemble belongs to region TR of the histogram $\protect\{ -110 <
E < -90\protect\}$. The insets show typical conformations in the folded
and unfolded regions.}
\label{fig:1}
\end{figure}

Thermodynamically, the folding transition in small proteins is analogous
to a first-order transition whereby two thermodynamic states
\cite{Jackson98} (folded and unfolded) are free energy minima while
intermediate states are unstable. The kinetic mechanism of transitions
from the unfolded state to the folded state is nucleation
\cite{Karpov96,Lifshits81,Shakhnovich97,Fersht97,Pande98}. Folding
nuclei can be defined as the minimal stable element of structure whose
existence results in subsequent rapid assembly of the native state. This
definition corresponds to a ``post-critical nucleus'' related to the
first stable structures that appear immediately after the transition
state is overcome \cite{Abkevich94}. The thermal probability of a
transition state conformation is low compared to the folded and unfolded
states, which are both accessible at the folding transition temperature
$T_f$ (see Fig.~\ref{fig:1}).

Kinetic analyses
\cite{Shakhnovich97,Pande98,Abkevich94,Shakhnovich98,Shakhnovich96,Mirny98,Gutin98}
for a number of lattice model chains of different lengths and degrees of
sequence design (optimization) point to a specific protein folding
nucleus scenario. Passing through the transition state with subsequent
rapid assembly of the native conformation requires the formation of some
(small) number of specific obligatory contacts (protein folding
nucleus). This result has been verified \cite{Shakhnovich96} for
sequences designed in the lattice model using different sets of
potentials, where it is shown that nucleus location was identical for
two different sequences designed with different potentials to fold into
the same structure of a lattice 48-mer. This finding and related results
\cite{Pande98,Klimov98} suggest that the folding nucleus location
depends more on the topology of the native structure than on a
particular sequence that folds into that structure.

The dominance of geometrical/topological factors in the determination of
the folding nucleus is a remarkable property that has evolutionary
implications (see below). It is important to understand the physical
origin of this property of folding proteins and assess its
generality. To this end, it is important to study other than lattice
models and other than Monte-Carlo dynamic algorithms. Here we employ the
discrete molecular dynamics (MD) simulation technique (the G\bo model
\cite{Taketomi75,Go81,Abe81} with the square-well potential of the
inter-residue interaction) to search for the nucleus in a continuous
off-lattice model \cite{Zhou97,Zhou97a,Dokholyan98}.

\bigskip
\bigskip
\noindent {\sf \large The transition region}

\noindent Our proposed method to search for a folding nucleus is based
on the observation \cite{Abkevich94} that equilibrium fluctuations
around the native conformation can be separated into ``local'' unfolding
(followed by immediate refolding) and ``global'' unfolding that leads to
a transition into an unfolded state and requires longer time to
refold. Local unfolding fluctuations are the ones that do not reach the
top of the free energy barrier and, hence, are committed to moving
quickly back to the native state. In contrast, global unfolding
fluctuations are the ones that overcome the barrier and are committed to
descend further to the unfolded state. Similarly, the fluctuations from
the unfolded state can be separated into those that descend back to the
unfolded state and those that result in productive folding. The
difference between the two modes of fluctuation is whether or not the
major free energy barrier is overcome. This means that the nucleation
contacts (i.~e. the ones that are formed on the ``top'' of the free
energy barrier as the chain passes it upon folding) should be identified
as contacts that are present in the ``maximally locally unfolded''
conformations but are lost in the globally unfolded conformations of
comparable energy.

\begin{figure}[htb]
\centerline{
\epsfxsize=8.0cm
\epsfbox{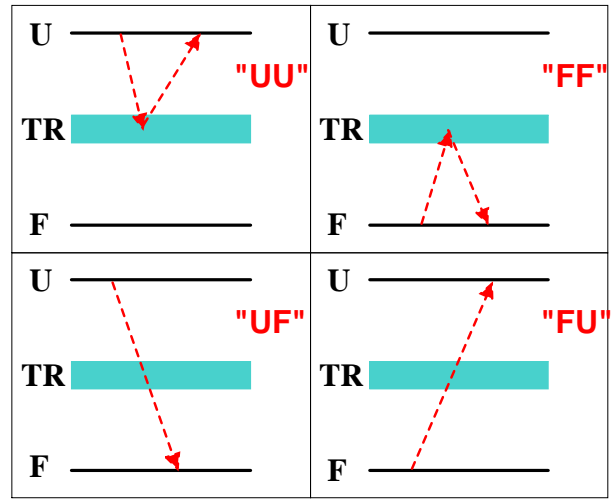} 
\vspace*{1.0cm}
}
\caption{The schematic definition of the four types of conformations:
{\sf FF}, {\sf UU}, {\sf UF}, and {\sf FU}.}
\label{fig:2}
\end{figure}

\begin{figure}[htb]
\centerline{
\epsfxsize=8.0cm
\epsfbox{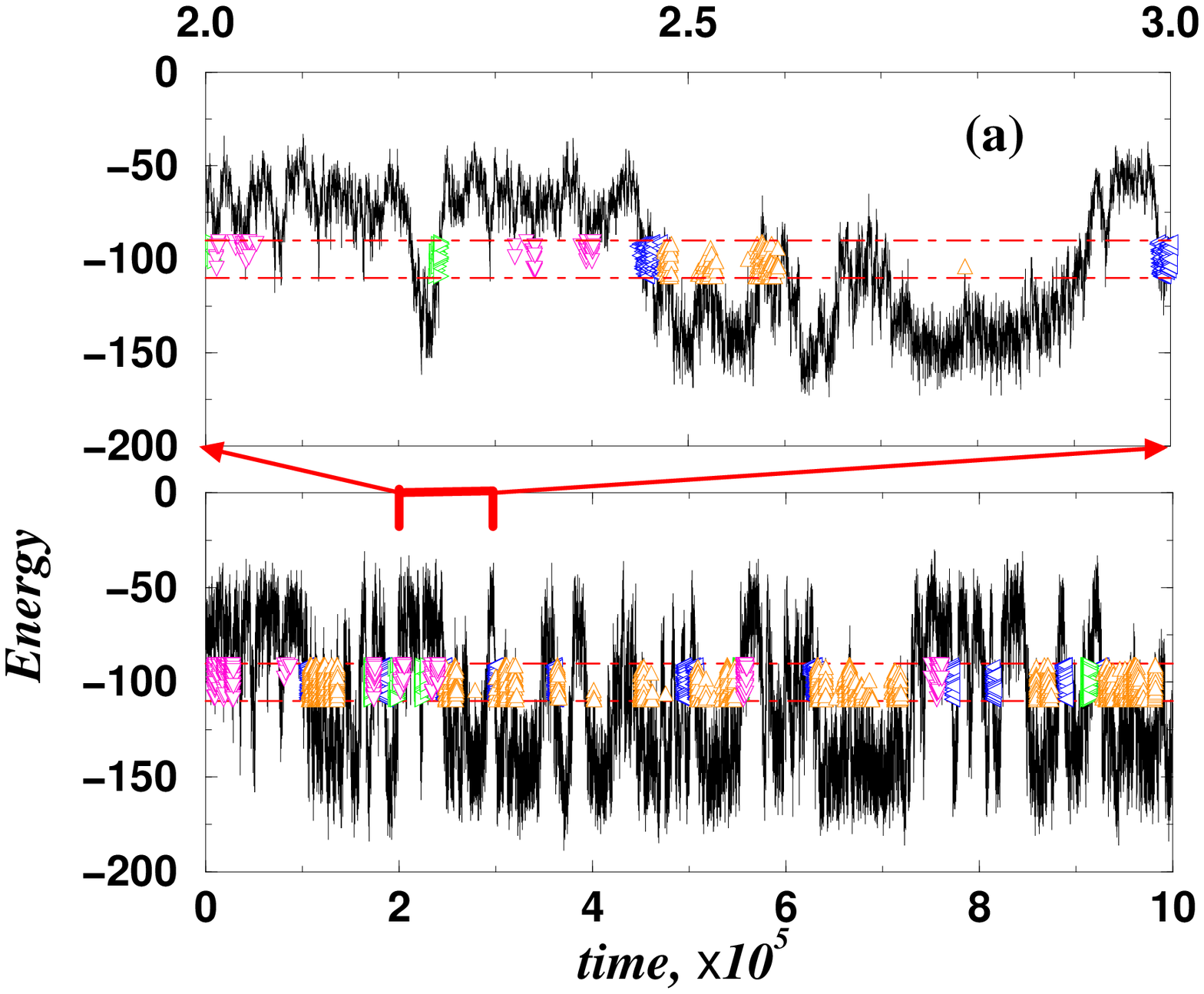}
\vspace*{1.0cm}
}
\newpage
\centerline{
\epsfxsize=8.0cm
\epsfbox{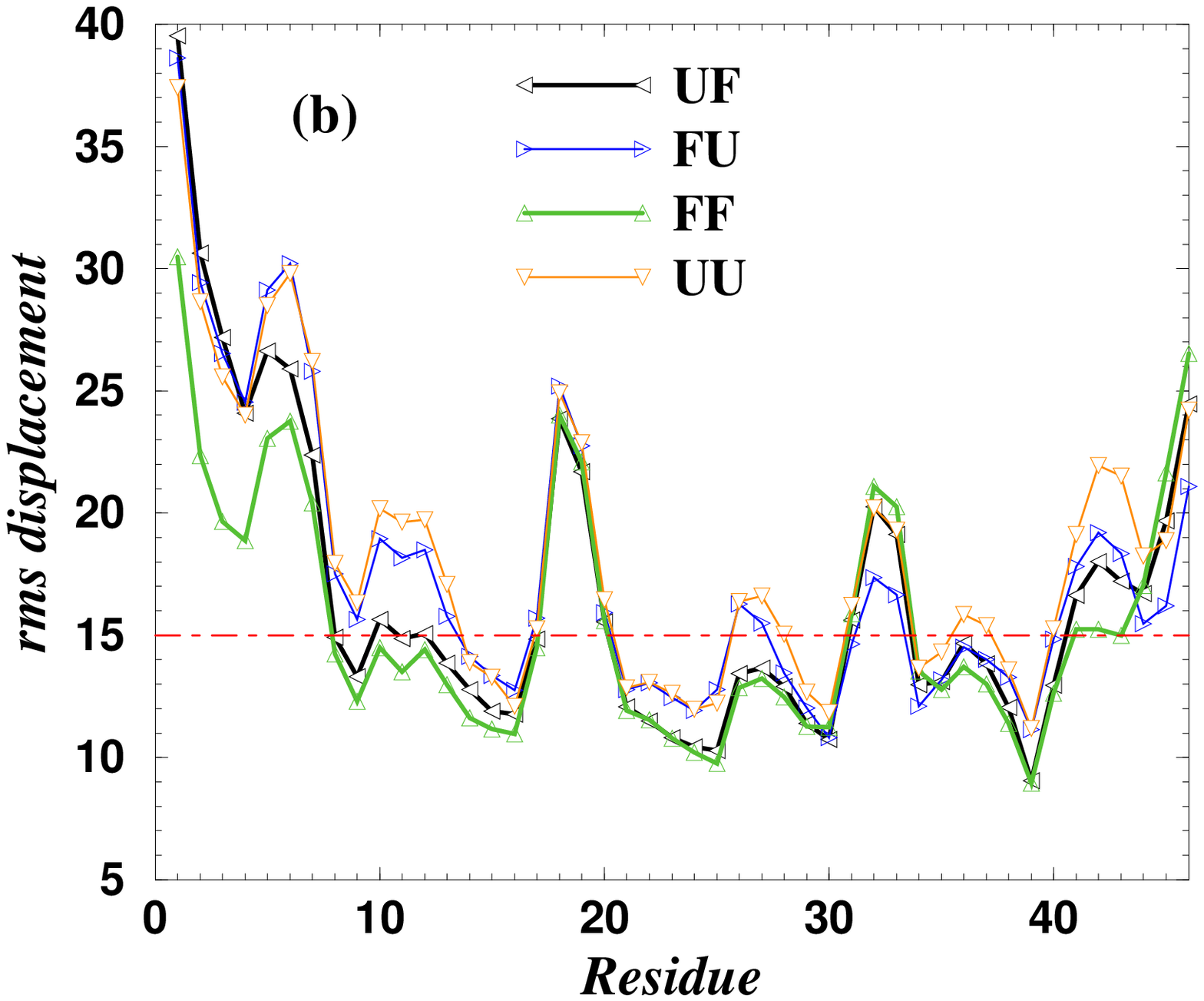}
\vspace*{1.0cm}
}
\centerline{
\epsfxsize=8.0cm
\epsfbox{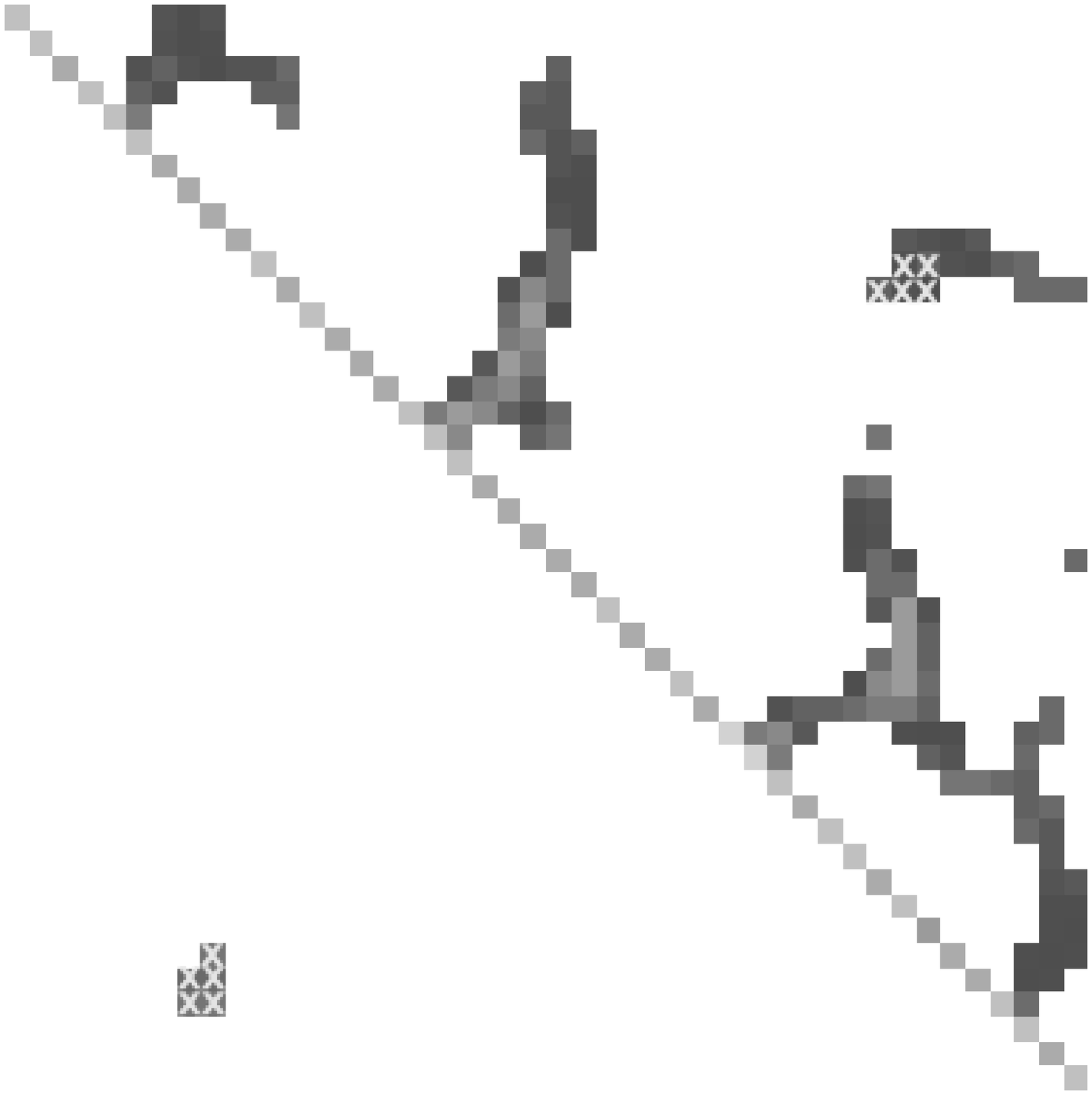}
\vspace*{1.0cm}
}
\centerline{
\epsfxsize=8.0cm
\epsfbox{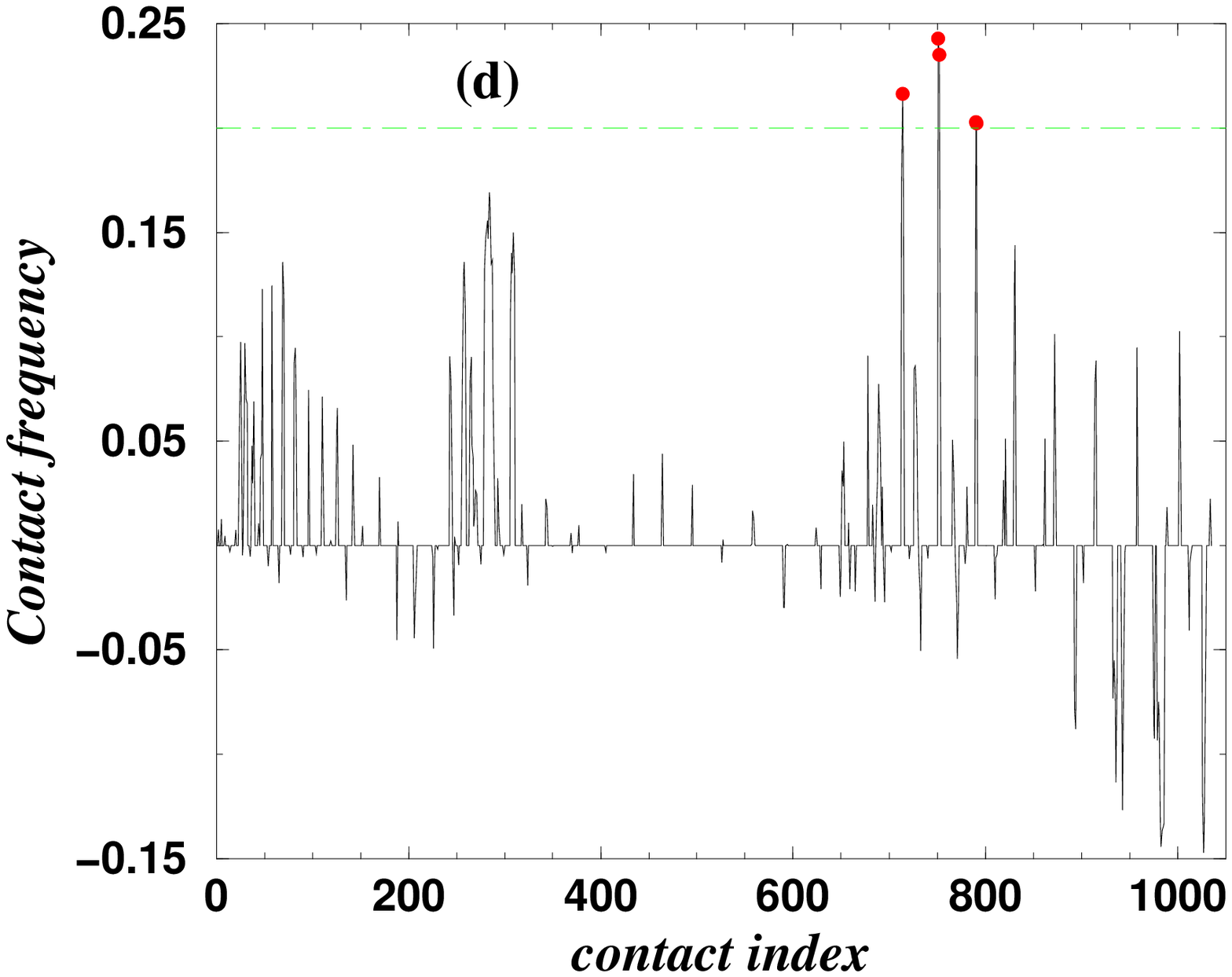}
\vspace*{1.0cm}
}
\caption{(a) The time evolution of the energy $E$ of the 46-mer
maintained at the folding temperature $T_f=1.44$. We focus on region TR:
$\protect\{ -110 < E < -90\protect\}$, which is denoted by two red
dash-dotted lines. The triangles with arrows pointing up, down, left and
right denote the conformations {\sf FF}, {\sf UU}, {\sf UF} and {\sf
FU}. The upper figure magnifies the region $2\cdot 10^5 \leq t \leq
3\cdot 10^5$. (b) The average rms displacement of each residue from its
native state for each of the four types of TR conformations: {\sf FF},
{\sf UU}, {\sf UF} and {\sf FU}. Note that there is a pronounced
difference in rms displacement of the residues from their native state
in the vicinity of the residues 10 and 40. The red dash-dotted line
indicates the breaking point of the native contacts, i.~e. when the rms
displacement $\sigma$ is approximately the size of the average relative
distance between pairs of residues, i.~e. $\sigma = (a_0+a_1)/2 \approx
15$ (see ``Methods'' section for the definition of $a_0$ and $a_1$). (c)
The contact map of the model protein. The darker the shade of grey, the
larger is the frequency of a contact. Above the diagonal of the square
matrix shows the {\protect\em native contacts} (see ``Methods'' section)
of the {\sf FF} conformations (if the {\protect\em native} contact
frequency is larger than 0.2).  Below the diagonal of the square matrix
shows the {\em difference between the frequencies of the native
contacts} in {\sf FF} and {\sf UU} conformations (if this difference is
larger than 0.2). Five contacts that persist in the {\sf FF}
conformations --- $(11, 39)$, $(10, 40)$, $(11, 40)$, $(10, 41)$, and
$(11, 41)$ --- are marked by crosses. The figure shows that
identification of the protein folding nucleus is facilitated by the
method used to construct the region of the matrix below the
diagonal. (d) The contact frequency difference of the {\sf FF}
conformations versus {\sf UU} conformations plotted versus the contact
index $k=i(i-1)/2 + j$, where $0\le j < i=1,2,\dots 45$. The
dashed-dotted line indicates the threshold value for the frequency
differences. The circles indicate the contacts whose frequency
differences are larger than the threshold. (Due to the scale used, only
four contacts can be seen in the figure.)}
\label{fig:3}
\end{figure}

Thus, in order to identify the folding nucleus, we study the
conformations of the 46-mer that appear in various kinds of folding
$\rightleftharpoons$ unfolding fluctuations. First, consider the time
behavior of the potential energy at $T_f$ (see Fig.~\ref{fig:3}a).  The
transition state conformations belong to the transition region TR from
the folded state to the unfolded state that lies in the energy range $\{
-110 <E < -90\}$ (see Fig.~\ref{fig:1}). Region TR corresponds to
the minimum of the histogram of the energy distribution. If we know the
past and the future of a certain conformation that belongs to the TR, we
can distinguish four types of such conformations (see Figs.~\ref{fig:2}
and~\ref{fig:3}a): {\em (i)} {\sf UU} conformations that originate in
and return to the unfolded region without ascending to the folded
region; {\em (ii)} {\sf FF} conformations that originate in and return
to the folded region without descending to the unfolded region; {\em
(iii)} {\sf UF} conformations that originate in the unfolded region and
descend to the folded region; and {\em (iv)} {\sf FU} conformations that
originate in the folded region and descend to the unfolded region.

If the nucleus exists, then the {\sf UF}, {\sf FU}, {\sf FF}, and {\sf
UU} conformations must have different properties depending on their
history. For example, the difference between {\sf UF}, {\sf FU}, {\sf
FF}, and {\sf UU} conformations is pronounced for the rms displacements
from the native state of the residues in the vicinity of the residues 10
and 40 and is illustrated in Fig.~\ref{fig:3}b. One difference between
the {\sf FF} conformations and {\sf UU} conformations is that the
protein folding nucleus is more likely to be retained in the {\sf FF}
conformations than in the {\sf UU} conformations. The contacts belonging
to the critical nucleus (``nucleation contacts'') start appearing in the
{\sf UF} conformations, and start disappearing in the {\sf FU}
conformations, so that the frequencies of nucleation contacts in {\sf
UF} and {\sf FU} conformations should be in between {\sf FF} and {\sf
UU}.

Our goal is to select the contacts that are crucial for the folding
$\rightleftharpoons$ unfolding transition. To this end we select the
contacts that appear much more often in the {\sf FF} conformations than
in the {\sf UU} conformations. We discover that if we set the threshold
for the difference in contact frequencies between {\sf FF} and {\sf UU}
conformations to be 0.2, then there are only five contacts that persist:
$(\mbox{residue~} 11, \mbox{residue~} 39)$, $(10, 40)$, $(11, 40)$,
$(10, 41)$, and $(11, 41)$ (see Fig.~\ref{fig:3}c,d). These contacts can
serve as evidence for the protein folding nucleus in the folding
$\rightleftharpoons$ unfolding transition in our model.

\begin{figure}[htb]
\centerline{
\epsfxsize=8.0cm
\epsfbox{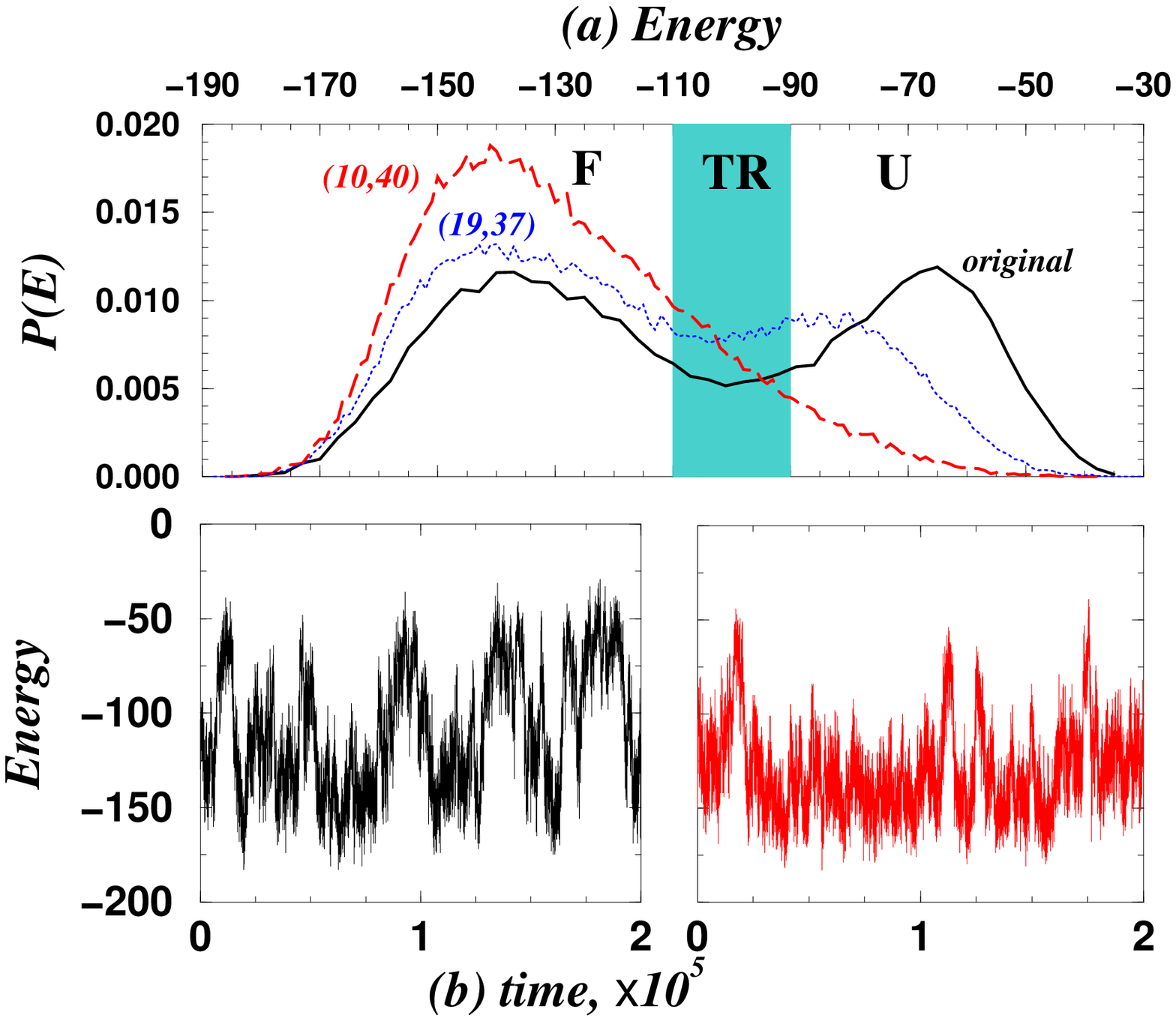} 
\vspace*{1.0cm}
}
\caption{(a) The probability distribution of the energy states $E$ of
the: {\em (i)} original 46-mer (at $T_f=1.44$), {\em (ii)} 46-mer (at
$T=1.46$) with a fixed contact belonging to the protein folding nucleus,
$(10,40)$, and {\em (iii)} 46-mer (at $T=1.46$) with fixed
randomly-chosen control contact $(19, 37)$, which does not belong to the
protein folding nucleus. Note that the probability of the unfolded state
of the 46-mer with a fixed contact belonging to the protein folding
nucleus, is suppressed compared to that of the original 46-mer. (b) The
time evolution of the energy $E$ of {\em (i)} original (left) and {\em
(ii)} fixed $(10,40)$ contact (right). Case {\em (iii)} fixed $(19, 37)$
contact is similar to {\em (i)}, so we do not show it. For case {\em
(i)}, the fluctuations are mostly between two extreme values of energy,
corresponding to the folded and unfolded states. In contrast, for case
{\em (ii)}, the fluctuations are mostly around one energy value,
corresponding to the folded state.}
\label{fig:4}
\end{figure}

Next, we demonstrate that these five selected contacts belong to the
protein folding nucleus. Suppose we fix just one of them, e.~g. $(10,
40)$, i.~e. we impose a covalent (``permanent'') link between residue
$10$ and residue $40$. If this contact belongs to the protein folding
nucleus, its fixation by a covalent bond would eliminate the barrier
between the folded and unfolded states, i.~e.  only the native basin of
attraction will remain. Hence, we hypothesize that the cooperative
transition between the unfolded and folded state will be eliminated and
the energy histogram (Fig.~\ref{fig:1}) should change qualitatively from
bimodal to unimodal. Our MD simulations support this hypothesis
(Fig.~\ref{fig:4}): fixation of only {\em one} nucleation contact, $(10,
40)$, gives rise to a qualitative change in the energy distribution from
bimodal to unimodal. Indeed, the probability to find an unfolded state
with a fixed link, $(10, 40)$, which belongs to the protein folding
nucleus, is drastically reduced compared to the probability of the
unfolded state of the original 46-mer, indicating the importance of the
selected contact $(10,40)$.

To provide a ``control'' that a {\em specific} contact plays such a
dramatic role in changing the character of the energy landscape, we fix
a randomly-chosen contact, $(19,37)$, which is not predicted by our
analysis, to belong to the critical nucleus. Our hypothesis predicts no
qualitative change in the energy distribution histogram since the
barrier, determined primarily by nucleation contacts, should not change
dramatically for this control. Fig.~\ref{fig:4} shows that this is
indeed the case. (The stability of the folded state is somewhat
increased for the control because any preformed native contacts decrease
the entropy of the unfolded state --- i.~e. they stabilize the folded
state).

We also find that for the {\sf UF} conformation that the rms
displacements of the residues from their native positions are smaller
than those for the {\sf FU} conformations (Fig.~\ref{fig:3}b). This
observation is consistent with the fact that the nucleation contacts are
formed first upon entering into productive folding and are destroyed
last upon unfolding.

\bigskip
\bigskip
\noindent {\sf \large Discussion}

\noindent Our main conclusion is that the existence of a few ($\approx
5$) specific contacts is signature of the transition state
conformations.  Those contacts can be defined as the protein folding
nucleus. Other contacts may also be present in transition state
conformations. However, they are optional and vary from conformation to
conformation, while nucleation contacts are present in transition state
conformations with high probability. Formation of nucleation contacts
can be considered as an obligatory step in the folding process: after
they are formed the major barrier is overcome and subsequent folding
proceeds ``downhill'' in the free energy landscape without encountering
any further major free energy barriers. This is illustrated by our
results that show that even one nucleation contact eliminates the free
energy barrier and, hence, leads to fast ``downhill'' motion to the
folded state. As a control our results show that fixation of an
arbitrary non-nucleation contact does not result in a similar effect.

The protein folding nucleus scenario of the transition state was
initially derived from Monte-Carlo studies of lattice models
\cite{Pande98,Abkevich94,Shakhnovich96} and was consistent with protein
engineering experiments with several small proteins
\cite{Itzhaki95,Martinez98}. Here, for the first time, we confirm this
scenario in the off-lattice MD simulations. The consistency between
conclusions made in different simulations
\cite{Pande98,Abkevich94,Shakhnovich96} and in experiments
\cite{Itzhaki95,Martinez98} is remarkable, and supports the possibility
that the protein folding nucleus formation is a generic scenario to
describe the protein folding transition state.

Our present study buttresses the point that the location of a protein
folding nucleus is determined by the geometry of the native state rather
than the energetics of interactions in the native state (the two factors
are not entirely independent, since native contacts must be generally
more stable to provide stability to the native conformation). In the
present study, we used the G\bo model (where all native contacts have
the same energy). Nonetheless, it turns out that some contacts
(nucleation) are ``more equal than others'' in terms of their role in
shaping the free energy landscape of the chain and determining folding
kinetics. This fact has implications for protein evolution, raising the
possibility that proteins that have similar structures but different
sequences may have similarly located protein folding nuclei. This
prediction was verified for SH3 domains \cite{Martinez98,Grantchanova98}
and for cold-shock proteins \cite{Perl98}. In terms of the evolutionary
selection of protein sequences, the robustness of the folding nucleus
suggests that any additional evolutionary pressure that controls the
folding rate may have been applied selectively to nucleus residues, so
that nucleation positions may have been under double (stability +
kinetics) pressure in all proteins that fold into a given
structure. Such additional evolutionary pressure has indeed been found
in the analysis of several protein superfamilies~\cite{Mirny98}.

\bigskip
\bigskip
\noindent {\sf \large Methods}

\noindent We study a ``beads on a string'' model of a protein. We model
the residues as hard spheres of unit mass. The potential of interaction
between residues is ``square-well''. We follow the G\bo model
\cite{Taketomi75,Go81,Abe81}, where the attractive potential between
residues is assigned to the pairs that are in contact in the native
state and repulsive potential is assigned to the pairs that are not in
contact in the native state. Thus, the potential energy is given by
\be
E = \frac{1}{2}\sum_{i,j=1}^{N}{U_{ij}}\, \,
\label{eq:U}
\ee
where $i$ and $j$ denote residues $i$ and $j$. $U_{ij}$ is the
matrix of pair interactions
\be
         U_{ij} = \left\{ \begin{array}{ll}
                   +\infty , & |r_i-r_j|\le a_0\\
                   -\Delta_{ij}\epsilon, &a_0<|r_i-r_j|\le a_1\\  
                   0, &  |r_i-r_j| > a_1\, .
                  \end{array} \label{eq:Uij}
\right.
\ee
Here, $a_0/2$ is the radius of the hard sphere, and $a_1/2$ is the radius
of the attractive sphere and $\epsilon$ sets the energy
scale. $||\Delta||$ is a matrix of contacts with elements
\be
\Delta_{ij} \equiv \left\{ \begin{array}{ll}
                   1, &  |r_i^{NS}-r_j^{NS}|\le a_1\\  
                  -1, &  |r_i^{NS}-r_j^{NS}| > a_1\, ,
                  \end{array} \label{eq:Delta}
\right.
\ee
where $r_i^{NS}$ is the position of the $i^{th}$ residue when the
protein is in the native conformation. Note that we penalize the
non-native contacts by imposing $\Delta_{ij}<0$. The parameters are
chosen as follows: $\epsilon=1$, $a_0=9.8$ and $a_1=19.5$. The covalent
bonds are also modeled by a square-well potential:
\be
      V_{i,i+1} = \left\{ \begin{array}{ll}
                   0, &  b_0 < |r_i-r_{i+1}| < b_1\\
                   +\infty, &  |r_i-r_{i+1}|\le b_0, \mbox{ or } 
                               |r_i-r_{i+1}|\ge b_1\, .  
                  \end{array} \label{eq:Vnn}
\right.
\ee

The values of $b_0=9.9$ and $b_1=10.1$ are chosen so that average
covalent bond length is equal to 10. The original configuration of the
protein ($N=46$ residues) was designed by collapse of a homopolymer at
low temperature \cite{Berriz97,Shakhnovich93,Abkevich96}. It contains
$n^* = 212$ native contacts, so the native state energy $E_{NS}
=-212$. The radius of gyration of the globule in the native state is
$R_G\approx 20$. The folding transition temperature $T_f=1.44$ is
determined by the location of the peak in the heat capacity dependence
on temperature.

Our simulations employ the discrete MD algorithm
\cite{Zhou97,Zhou97a,Dokholyan98}. To control the temperature of the
protein we introduce 954 particles that interact with the protein and
with each other via hard-core collisions and so serve as a ``heat
bath''. Thus, by changing the kinetic energy of those heat bath
particles we are able to control the temperature of the environment. The
heat bath particles are hard spheres of the same radii as the chain
residues and have unit mass. Temperature is measured in units of
$\epsilon/k_B$. The variable time step is defined by the shortest time
between two consecutive collisions.

\bigskip
\bigskip
\noindent {\sf Acknowledgments} 

\noindent We thank R. S. Dokholyan for careful reading of the
manuscript. NVD is supported by NIH NRSA molecular biophysics
predoctoral traineeship (GM08291-09). EIS is supported by NIH grant
RO1-52126. The Center for Polymer Studies acknowledges the support of
the NSF.

\bigskip
\bigskip
\noindent {\large \sf
Nikolay V. Dokholyan$^1$, Sergey V. Buldyrev$^1$, H. Eugene Stanley$^1$
and Eugene I. Shakhnovich$^2$}

\bigskip
\bigskip
\noindent $^1$Center for Polymer Studies, Physics Department, Boston
University, Boston, MA 02215, USA

\bigskip
\noindent $^2$Department of Chemistry, Harvard University, 12 Oxford
Street, Cambridge, MA 02138, USA

\bigskip
\bigskip
\noindent Correspondence should be addressed to N.V.D. {\sl email:
dokh@bu.edu}

%\section{Legends}

\end{document}